\documentclass[aip,reprint,graphicx]{revtex4-1}
\usepackage{CJK}
\usepackage{graphicx}
\usepackage{dcolumn}
\usepackage{bm}
\usepackage{verbatim}

\begin{document}
\begin{CJK*}{GB}{}

\title{A Rayleigh-Brillouin scattering spectrometer for ultraviolet wavelengths} 



\author{Ziyu Gu\CJKfamily{gbsn}(¹È×ÓÓð)}
\affiliation{Institute for Lasers, Life and BioPhotonics, VU University Amsterdam,
De Boelelaan 1081, 1081 HV  Amsterdam, The Netherlands}

\author{M. Ofelia Vieitez}
\affiliation{Institute for Lasers, Life and BioPhotonics, VU University Amsterdam,
De Boelelaan 1081, 1081 HV  Amsterdam, The Netherlands}

\author{Eric-Jan van Duijn}
\affiliation{Institute for Lasers, Life and BioPhotonics, VU University Amsterdam,
De Boelelaan 1081, 1081 HV  Amsterdam, The Netherlands}

\author{Wim Ubachs}
\email{w.m.g.ubachs@vu.nl}
\affiliation{Institute for Lasers, Life and BioPhotonics, VU University Amsterdam,
De Boelelaan 1081, 1081 HV  Amsterdam, The Netherlands}

\date{\today}

\begin{abstract}
A spectrometer for the measurement of spontaneous Rayleigh-Brillouin scattering line profiles at ultraviolet wavelengths from gas phase molecules has been developed, employing a high-power frequency-stabilized UV-laser with narrow bandwidth (2 MHz). The UV light from a frequency-doubled titanium:sapphire laser is further amplified in an enhancement cavity, delivering a 5 Watt UV-beam propagating through the interaction region inside a scattering cell. The design of the RB-scattering cell allows for measurements at gas pressures in the range $0-4$ bar and at stably controlled temperatures from $-$30$^{\circ}$C to 70$^{\circ}$C. A scannable Fabry-Perot analyzer with instrument resolution of 232 MHz probes the Rayleigh-Brillouin profiles. Measurements on N$_2$ and SF$_6$ gases demonstrate the high signal-to-noise ratio achievable with the instrument, at the 1\% level at the peak amplitude of the scattering profile.
\end{abstract}


\maketitle 

\end{CJK*}
\section{Introduction}
\label{intro}
Rayleigh-Brillouin (RB) light scattering is a powerful method to investigate intrinsic thermodynamic material properties, such as thermal diffusivity, speed of sound, heat capacity ratios, and relaxation times of various dynamical processes occurring in media. Immediately after the invention of the laser as a source of narrow bandwidth radiation in the 1960s, techniques were developed to measure the characteristic scattering profiles, resolving the Brillouin doublet peaks shifted from a central elastic Rayleigh peak, first in the liquid and solid phase\cite{Benedek1964}$^,$\cite{Chiao1964} and subsequently in the gas phase\cite{Greytak1966}. During the 1970s detailed studies on RB-scattering in gas phase media were performed, in particular on molecular hydrogen~\cite{Hara1971}, on molecular nitrogen\cite{Sandoval1976}, on various polyatomic gases exhibiting internal relaxation~\cite{Lao1976}, and on deriving scaling laws for the noble gases~\cite{GhaemMaghami1980}. Over the years various formalisms were derived to describe the spectral scattering profiles based on density fluctuations in the hydrodynamic regime~\cite{Mountain1966} and in the kinetic regime\cite{Boley1972}$^,$\cite{Tenti1974}. The latter models, that have become known as the Tenti-models, have been most succesfull in describing RB-scattering over a wide range of conditions, including the transition from the kinetic to the hydrodynamic regime. In particular, in recent studies the 6-component version of the Tenti-model\cite{Tenti1974} (Tenti S6) was found to accurately describe the RB-scattering profile in various atomic and molecular gases~\cite{Vieitez2010} and in air~\cite{Witschas2010}.

The Rayleigh-Brillouin scattering profile can be measured by different means. The direct method analyzes the scattered light via a Fabry-Perot interferometer, as was pursued in the early studies\cite{Greytak1966}$^,$\cite{Lao1976}$^,$\cite{Letamendia1981}. Later various forms of optical beating were pursued~\cite{Eden1974,Matsuoka1993}, including superheterodyne techniques making use of frequency tunable lasers~\cite{Tanaka2002}. While most studies have been directed towards measuring the spectral profiles in spontaneous RB-scattering, in the last decade also methods were developed for the investigation of coherent RB-scattering~\cite{Pan2002,Pan2004,Meijer2010}.

In this paper, we describe an apparatus to accurately measure spectral profiles of Rayleigh-Brillouin scattering in molecular gases at ultraviolet wavelengths. The apparatus employs a high-intensity continuous-wave narrowband laser source in the ultraviolet range, extended with an enhancement cavity yielding 5 W of scattering power to record  the RB-profiles at 1\% peak-intensity fluctuations even at sub-atmospheric pressures. Such accurate assessment of the scattering profiles of air at atmospheric and sub-atmospheric pressures is required for implementation in modern Doppler-wind remote sensing applications as envisioned for the ADM-Aeolus satellite mission of the European Space Agency (ESA)~\cite{ESA-SP-1311}.

\section{Experimental Apparatus}

\begin{figure*}
\includegraphics[width=\linewidth]{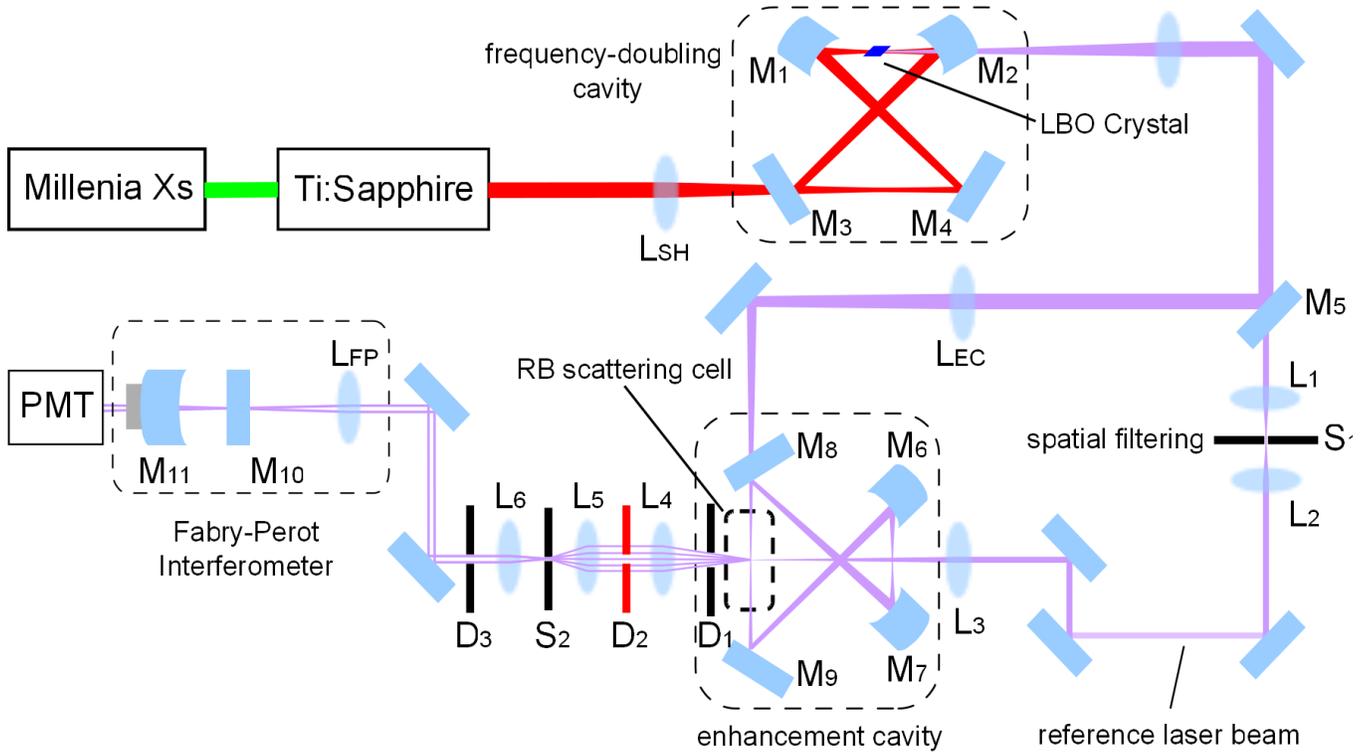}
\caption{\label{fig:whole_setup} Schematic of the experimental apparatus. The Ti:Sa laser, pumped by a 10 W Millennia Xs pump laser, yields after frequency doubling (in an LBO crystal) in an external enhancement cavity a cw power of 500 mW at a UV-wavelength of 366.8 nm. The UV laser beam is then directed into a second enhancement cavity for amplification by a factor of 10. The RB-scattering cell is placed in the intra-cavity focus of the UV beam to ensure a maximum scattering intensity. Scattered light is collected at $90^{\circ}$ with respect to the beam direction. The geometrically filtered scattered light is directed to the Fabry-Perot Interferometer (FPI), which has a $232\pm4$ MHz instrumental linewidth. The photons transmitted by the FPI are detected by a photo-multiplier tube (PMT). Note that a small fraction of the UV-light transmitted through mirror M$_5$ is used as a reference beam for aligning beam paths and for characterizing the detecting system. Mirrors, lenses and diaphragms are indicated with M$_i$, L$_i$ and D$_i$.}
\end{figure*}

The experimental apparatus for measuring Rayleigh-Brillouin scattering is sketched in Fig.~\ref{fig:whole_setup}. It consists of a narrowband tunable laser source with an external frequency-doubling cavity for the production of UV-light, an RB-scattering cell mounted inside an enhancement cavity for increasing the effective scattering power, and a Fabry-Perot Interferometer (FPI) for analyzing the spectral profile of the scattered light. The choice was made to detect at a $90^{\circ}$ scattering angle. The apparatus and the comprising units are further described in the following subsections.

\subsection{Characteristics of the laser source}\label{sec:laser}

The laser system is based on a continuous-wave (cw) Titanium:Sapphire (Ti:Sa) ring laser (Coherent 899-21), pumped by a frequency-doubled Nd:YVO$_4$ pump laser (Spectra-Physics Millennia Xs) with 10 W power at 532 nm. The wavelength of the Ti:Sa laser is tunable in the range 690 to 1100 nm, with a maximum output power more than 2 W near 800 nm. While at its second harmonic (corresponding to $400$ nm) in excess of 1 W could be produced, the actual wavelength for RB-scattering was chosen deeper into the UV, compromising between a high scattering cross section and high achievable UV-powers; though in many remote sensing LIDAR applications $355$ nm is the wavelength of choice, most of the experiments were carried out at 366.8 nm, where the UV yield of the laser system is much higher. For this setting the Ti:Sa laser produces $1.5$ W at $733.6$ nm at $1$ MHz bandwidth.

An external frequency-doubling cavity, with a brewster-cut lithium-triborate (LiB$_3$O$_5$; LBO) nonlinear crystal mounted in the focus was employed for efficient production of UV light. Methods of impedance matching, phase matching, mode matching and locking the cavity by a H\"{a}nsch-Couillaud scheme were detailed in a previous report from our laboratory~\cite{Koelemeij2005}. This unit delivers an output power of more than $500$ mW at $366.8$ nm, with a bandwidth estimated at 2 MHz. This narrow bandwidth is the reason for choosing a CW laser source. While pulsed lasers might have some advantages in noise rejection and detection, a bandwidth of 2 MHz cannot be achieved with pulsed lasers suffering from Fourier limitations.

Locking the Ti:Sa laser to a reference cavity ensures a frequency drift limited to $18$ MHz per hour, as measured by a wave meter (ATOS Lambdameter), which itself exhibits a drift limited to $10$ MHz/hour. Therefore, the actual drift of the frequency is expected to be less than $36$ MHz/hour at UV wavelengths.

\subsection{Scattering Cell and Enhancement Cavity}\label{sec:scattering cell and enhancement cavity}

\begin{figure*}
\includegraphics{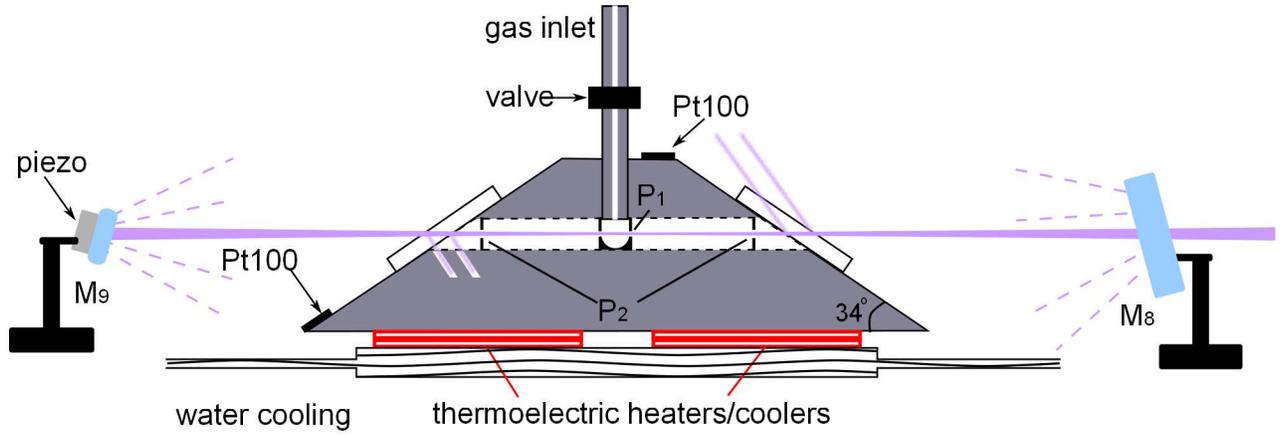}
\caption{\label{fig:scattering_cell_temperature_control} Side view of the RB-scattering cell. The UV beam enters the enhancement cavity through mirror M$_8$. P$_1$ and P$_2$ are pinholes for blocking photons scattered from the mirrors and the windows. Pt100 elements are temperature sensors.}
\end{figure*}

To achieve a high scattering signal, amplification of the UV laser beam is accomplished by a second enhancement cavity, similarly built as the frequency-doubling cavity: a flat input coupler M$_8$, a small flat mirror M$_9$ attached to a piezo tube for H\"{a}nsch-Couillaud locking, and two curved mirrors M$_6$ and M$_7$ with $-75$ mm radius-of-curvature. The center of the RB scattering cell is placed at the mid-point of M$_8$ and M$_9$, where the UV laser beam is focused to $\sim$ 200 $\mu$m by the lens L$_{\text{EC}}$ for mode matching of the in-coupling beam to the cavity. This enhancement cavity yields a power enhancement of 10 times at 366.8 nm, thus delivering a UV-light intensity of 5 W in the scattering volume.

Fig.~\ref{fig:scattering_cell_temperature_control} displays a side view of the RB-scattering cell mounted inside the enhancement cavity. The trapezoidal shape of the cell, with a 34$^{\circ}$ leg-to-base angle, maintains the windows tilted at Brewster's angles for 366.8 nm to reduce reflection losses and to introduce a polarization-dependent element required for H\"{a}nsch-Couillaud locking. The RB scattering signal is collected from the center of the cell, with a direction perpendicular to the incident beam. Two additional windows, with anti-reflection coatings on both sides, are mounted at the front and rear surfaces of the cell, in order to transmit the alignment laser beam and the Rayleigh-Brillouin scattered light in the direction of the detector.

The gas is let in and pumped out through the valve mounted on the top of the RB-scattering cell. The windows, sealed with viton O-rings, allow for a pressure variation from 0 bar to 4 bar. Four Peltier elements (indicated in red), disciplined by a temperature controlled water cooling system, can be used both as coolers and heaters and allow for a temperature variation of the gas sample from $-$30$^{\circ}$C to 70$^{\circ}$C. The temperature of the cell and the gas contained inside is measured by two Pt-100 elements stuck to the top right and bottom left corners of the cell. At two extreme conditions, namely $-30^{\circ}$C and 70$^{\circ}$C, these two elements display less than 0.5$^{\circ}$C reading difference, indicating that the temperature distribution is homogeneous. The cell itself, machined from solid aluminum, is capable of maintaining the temperature of the sample gas constant over a measuring period of typically 3 hours.

Several measures are taken to reduce stray light reaching the detector to a minimum level. Two pairs of pinholes, P$_1$ with 1 mm in diameter and P$_2$ with 1.5 mm in diameter, are placed along the beam path to filter out the UV light scattered from the cavity mirrors M$_8$ and M$_9$ as well as from the surfaces of the Brewster-cut entrance and exit windows for the UV-laser beam (see Fig.~\ref{fig:scattering_cell_temperature_control}).
The remaining reflections from the two sides of the exit window are captured and absorbed by two light traps, drilled inside the bottom of the cell and painted black inside. By aligning the incident and the reflected beams to pass exactly through the center of each pinhole, it is ensured that only the light scattered by the gaseous molecules will be detected.

\subsection{Light collection and alignment}\label{sec:Light_collection}

Scattered light is collected from the scattering volume inside the cell by a sequence of optical elements, which have the function to select a narrow opening angle for the scattered light, therewith defining the scattering geometry, to convert the scattered light into a collimated beam that can be accepted by the FPI analyzer, and to reduce the amount of stray light reaching the detector.

A low power auxiliary UV laser beam, leaking through mirror M$_5$ (in Fig.~\ref{fig:whole_setup}) is used to adjust the beam cleaning optics and to align and characterize the FPI. This reference laser beam is aligned to exactly cross the RB-scattering interaction volume and is subsequently used to fine-adjust the lenses L$_4$ to L$_6$ and to center the diaphragms D$_1$ and D$_3$. The light emerging from the interaction volume is mode-cleaned by a spatial filter S$_2$ (diameter 50 $\mu$m) in combination with two confocal lenses L$_5$ and L$_6$ (both $f=50$ mm). The collimated output of the light cleaning section is further narrowed by diaphragm D$_3$ ($2.5$ mm diameter) and is coupled by lens L$_{\text{FP}}$ to the FPI. This sequence of optics serves to match the acceptance mode profile of the FPI, while the narrow acceptance toward the FPI effectively reduces stray light originating from other locations than the scattering center.

\begin{figure}
\includegraphics{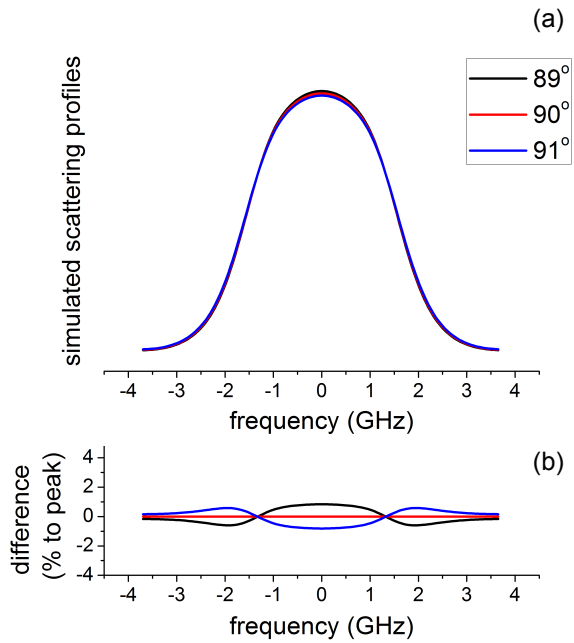}
\caption{\label{fig:scattering_angle_comparison} (a) Rayleigh-Brillouin scattering profiles for 1 bar N$_2$ at $24^{\circ}$C simulated with the Tenti S6 model for scattering angles of 89$^{\circ}$ (black), 90$^{\circ}$ (red) and 91$^{\circ}$ (blue). All the spectra are normalized to area unity for comparison. (b) Calculated deviations of RB-scattering profiles measured at 89$^{\circ}$ and 91$^{\circ}$ from that measured at 90$^{\circ}$. The deviation are shown in percentages of the peak amplitude at 90$^{\circ}$.}
\end{figure}

The opening angle of the RB-scattering geometry is controlled by the diaphragm D$_1$. Its diameter is kept at 0.8 $\pm$ 0.2 mm, while D$_1$ is placed 31 $\pm$ 1 mm away from the scattering center, thus yielding an opening angle of 0.7 $\pm$ 0.2$^{\circ}$. To assess the effect of angular alignment of the scattering geometry RB-scattering profiles are simulated by the Tenti-S6 model. RB profiles have been calculated for N$_2$ at 1 bar and 24$^{\circ}$C, assuming the scattering to 89$^{\circ}$, 90$^{\circ}$ and 91$^{\circ}$, respectively. The scattering profiles are normalized and compared in Fig.~\ref{fig:scattering_angle_comparison} (a). Deviations between these spectra, as plotted in Fig.~\ref{fig:scattering_angle_comparison} (b), indicate that for near-perpendicular scattering geometries amplitude deviations of up to 1 percent may occur as a result of a wrong estimation of the scattering angle by 1$^{\circ}$.

\subsection{Fabry-Perot Interferometer}\label{sec:FPI}

\begin{figure}
\includegraphics{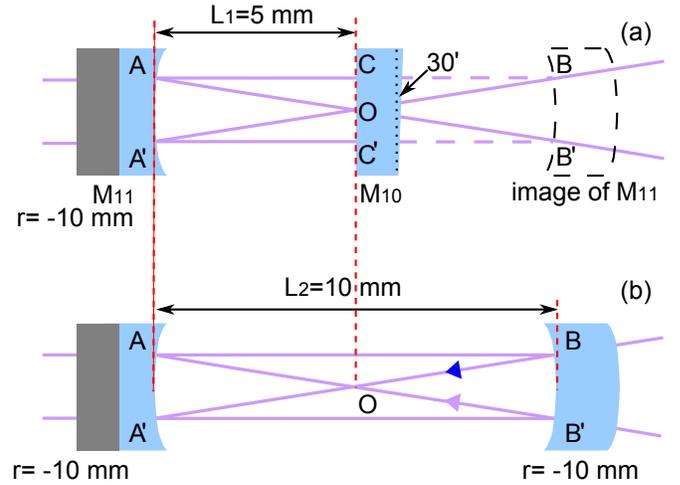}
\caption{\label{fig:Fabry_Perot} (a) The design of the Fabry-Perot Interferometer. The in-coupling mirror M$_{10}$ is flat and wedged by 30'. The out-coupling mirror M$_{11}$ is a concave mirror, with radius of curvature being $-10$ mm. The distance between these two mirrors is adjusted to be 5 mm. Thus the focal point O of M$_{11}$ is on the left surface of M$_{10}$. The out-coupler is mounted on a piezo for scanning. (b) A confocal Fabry-Perot Interferometer consisting of two concave mirrors both with the same radius of curvature as M$_{11}$. This FPI is shown to explain the fourfold mode pattern in the plano-concave interferometer.}
\end{figure}

To measure the RB scattering profile, a Fabry-Perot Interferometer, consisting of a flat in-coupling mirror and a concave out-coupling mirror, has been constructed. The rear sides of both mirrors (M$_{10}$ and M$_{11}$) are anti-reflection coated for 366.8 nm to minimize losses. The 30' wedge on the in-coupler (M$_{10}$) serves to suppress mode structure arising from the reflections between its surfaces. Compromised by the coupling efficiency and the finesse, the reflectivity of the out-coupling mirror (M$_{11}$) is chosen to be $R=99.0 \pm 0.2$\% for 366.8 nm. For impedance matching the reflectivity of the in-coupling mirror (M$_{10}$) is chosen to be $R=98.0 \pm 0.2$\%. With these choices and settings the in-coupling efficiency is around 75\% when the FPI is on resonance.

This plano-concave interferometer is in fact a folded spherical FPI sharing the advantages of a high light gathering power, proportional to the resolving power, and its insensitivity to small variations of incident angle~\cite{vaughan1989fabry}. In Fig.~\ref{fig:Fabry_Perot} the folded plano-concave FPI of length $L_1$ (in panel (a)) is compared with the fully confocal FPI of length $L_2=2L_1$. By carefully setting the mirror spacing to 5 mm, the focal point of M$_{11}$, which has a radius of curvature $r$ of $-$10 mm, is pointing on the left surface of M$_{10}$. The in-coupling lens L$_{\mathrm{FP}}$, with a focal length of $f=50$ mm, is positioned such that its focal point coincides with that of M$_{11}$. Hence, the collimated light beam incident from the light collection section, is mode matched to couple into the FPI. The out-coupling mirror M$_{11}$ is mounted on a piezo tube which serves as the scanning element to retrieve the RB-scattering spectra.

\subsubsection{Free Spectral Range}
\begin{figure}
\includegraphics{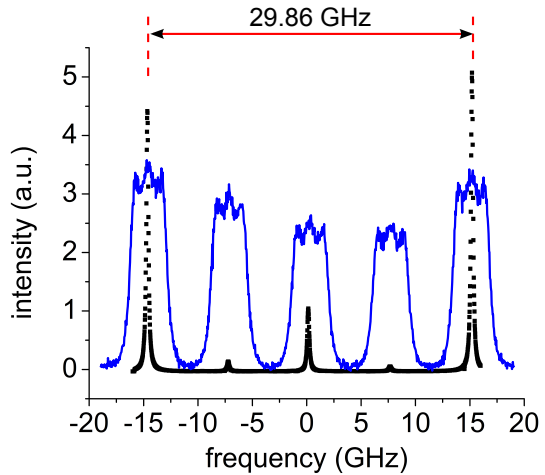}%
\caption{\label{fig:free_spectra_range} The FPI transmission curve probing the output of the coherent UV-laser is indicated by the (black) dots. The full (blue) curves represent the FPI-transmission of the Rayleigh-Brillouin  scattered light. Since the scattered light is not spatially coherent nor fully paraxial, the amplitude of all modes connected to the four-transit path interference appear equally strong.}
\end{figure}

For a FPI with two plane mirrors, the free spectral range (FSR) is $c/2nL$, resulting from the self interference of light beams between round trips ($2nL$), where $n$ is the refractive index of the material in between mirrors, $L$ the distance between the mirrors, and $c$ the speed of light in vacuum. The FPI is operated under ambient conditions. For a confocal FPI with mirror separation $L$ equal to the common radius of curvature $r$ of both mirrors, it can be shown in a ray-tracing analysis that optical rays retrace their paths after four successive reflections~\cite{vaughan1989fabry}. This results in a FSR of $c/4nL$ for the confocal geometry as displayed in Fig.~\ref{fig:free_spectra_range}(b). Therefore, the effective FSR of our plano-concave FPI of length $L_1$, mimicked by a spherical FPI of length $L_2$ is
\begin{equation}
\frac{c}{4nL_2}=\frac{c}{8nL_1}\approx 7.5 ~\text{GHz}
\end{equation}

As discussed in Ref.~\onlinecite{vaughan1989fabry}, however, if the incident beam is spatially coherent, additional interferences between different light rays in the confocal FPI can occur. For instance, after two reflections, the ray indicated in blue arrow in Fig.~\ref{fig:free_spectra_range}(b) will follow the same path as the ray indicated in purple arrow, and interference can occur. This phenomenon effectively enlarges the FSR by a factor of 2. However, this does not occur for incoherent light as produced in RB-scattering. In addition, if the incident beam is extremely paraxial, the plano-concave FPI almost works as a plane FPI, yielding an effective FSR of $c/2nL_1$, corresponding to $\sim 30$ GHz.

These phenomena are experimentally demonstrated by recordings of FPI transmission fringes by (i) making use of the spatially coherent reference laser beam leaking through mirror M$_5$ and aligned paraxially through the scattering cell, and (ii) by monitoring the transmission fringes of RB incoherently scattered light. Results are shown in Fig.~\ref{fig:free_spectra_range}. The FPI transmission fringes for the reference laser are monitored by continuously scanning its frequency, while keeping the FPI-mirrors at fixed distance; here the frequency separations are calibrated with an ATOS-wavelength meter, having a relative accuracy better than 50 MHz. The FPI transmission pattern shows the major fringes with separations of $\sim 30$ GHz corresponding to the paraxial alignment of the reference beam (FSR=$c/2nL_1$). The center fringe at 0 GHz in Fig.~\ref{fig:free_spectra_range} results from the wave-vector mismatch of the additional interferences in the plano-concave FPI. The two smaller peaks are due to the incomplete coherence of the reference beam~\cite{vaughan1989fabry}. The measurements of transmission fringes for RB-scattered light were performed by keeping the laser frequency fixed and by scanning the piezo-voltage on the FPI, and performing the calibration by interpolation (see below in section~\ref{sec:data_processing}). These spectra show all four modes, spaced by $\sim 7.5$ GHz, as expected for incoherent light.

\subsubsection{Instrument Function}
\begin{figure}
\includegraphics{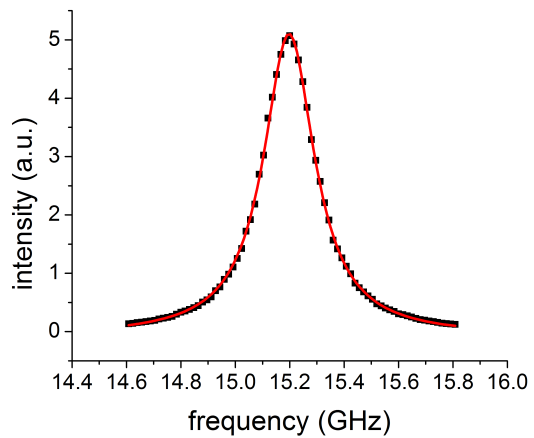}%
\caption{\label{fig:instrument_function} The transmission intensity of the FPI with respect to the frequency change of the reference laser. The Lorentzian linewidth of the transmission curve indicates the width of the instrument function, being 228.8 $\pm$ 1.4 $\text{MHz}$ in this specific measurement.}
\end{figure}

The instrument function of the FPI is characterized in the same way as the FSR in a continuous scan of the UV-laser. Fig.~\ref{fig:instrument_function} shows a typical measurement of the FPI transmission function. The experimental data (black dots) are fitted to a Lorentzian profile function (indicated in red curve), delivering a line width of $228.8 \pm 1.4$ MHz for this specific measurement. Reproducibility tests yield a mean value of $232\pm4$ MHz for the fringe width, which then determines the resolution $\Delta_{\mathrm{I}}$ of the FPI Rayleigh-Brillouin spectrum analyzer.  The laser bandwidth (2 MHz) is so small that it does not effectively contribute to the instrument linewidth of the FPI.

\section{Experimental Methods \& Results}

\subsection{Data Processing Procedure}\label{sec:data_processing}

\begin{figure}
\includegraphics{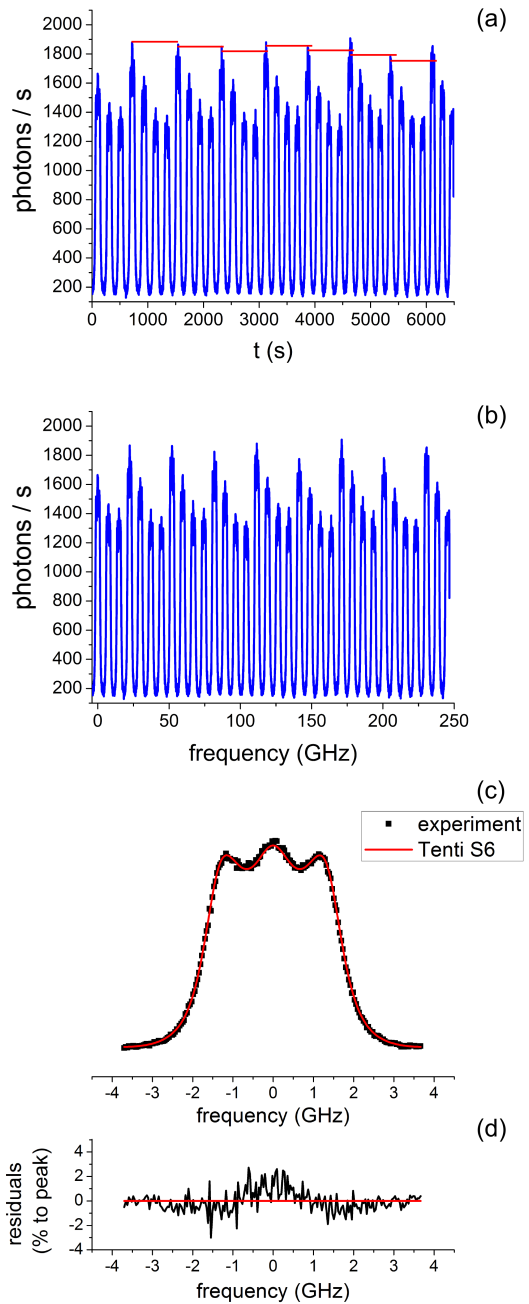}
\caption{\label{fig:data_processing1} (a) The recorded RB scattering intensity from N$_2$ at 3 bar and 23.6$^{\circ}$C for 90$^{\circ}$ scattering angle. The exposure duration of the photon multiplier tube is 1 s for each data point. (b) The RB scattering intensity on the horizontal scale converted to a frequency axis. (c) The final RB scattering profile (indicated in black dots), averaged from the spectra in (b) and normalized in unit area. The red curve is the calculated theoretical Tenti S6 model for comparison. Fig. (d) shows the difference between the measurement and the Tenti S6 model, in percentage of the maximum amplitude.}
\end{figure}

Rayleigh-Brillouin scattering profiles are measured by scanning the length of the FPI, while keeping the laser frequency fixed. An amplified computer-controlled voltage is applied to the piezo tube attached to the FPI out-coupling mirror M$_{11}$. Typical scans cover a voltage increasing from 0 V to 500 V at step sizes of 0.05 V. The piezo retracts over a maximal distance of 4.5 $\mu$m, resulting in frequency scans  of tens of effective FSR's. Data acquisition proceeds by detecting the transmitted light by a photo-multiplier tube (PMT) at typical exposure times of 1 s per step.

Fig.~\ref{fig:data_processing1} illustrates the measurement and calibration procedures for determining the RB scattering profiles. Part (a) shows an example measurement of an RB scattering time trace for N$_2$ at 3 bar, recorded at 23.6$^{\circ}$C for 1.8 hours (6500 seconds). Along the vertical axis the absolute number of photon counts detected per second is plotted. The horizontal (red) bars in Fig~\ref{fig:data_processing1}(a) connecting the main transmission fringes separated by 30 GHz, show a non-linearity along the time axis. There are several causes for this phenomenon: (i) the nonlinear conversion of piezo voltage into distance in the FPI; (ii) a temperature-induced drift of the FPI; (iii) frequency drift of the laser. The first effect gives the dominant contribution. The measuring time traces are linearized and converted to a frequency scale by fitting the transmission peaks to a series of Lorentzians, and subsequently linearizing the horizontal scale by matching the peak separations to the measured FSR. This results in the series of RB profiles along a frequency axis as displayed in Fig.~\ref{fig:data_processing1}(b).
This series of RB-profiles are cut at the midpoints between the transmission peaks into individual RB-profiles. The individual spectra are finally added and normalized to area unity to yield a final RB-scattering profile as shown in Fig~\ref{fig:data_processing1}(c).

The experimental profile may be compared to a theoretical description in terms of the Tenti S6 model~\cite{Vieitez2010,Witschas2010}. The calculated S6 curves are convolved with the measured instrument function of width 232 MHz; the fact that the series of overlapping RB-profiles measured by the FPI never reaches the zero level is accounted for (see also Ref.~[\onlinecite{Vieitez2010}]). A result of the convolved S6 model is shown as the full (red) line in Fig~\ref{fig:data_processing1}(c).

In the final panel (Fig.~\ref{fig:data_processing1}(d)) the residuals between the Tenti S6 model and experiment are shown on a percentage scale (of the full amplitude). This final result indicates that the difference between measured RB-profile and the Tenti S6-theory are at the 1\% level of the maximum amplitude, with some outliers to the 2\% level. The result of Fig.~\ref{fig:data_processing1}(d) also demonstrates that the rms measurement noise in the neighborhood of the RB-peak amplitude is at the 1\% level.

\subsection{Temperature-dependent Rayleigh-Brillouin scattering}
\begin{figure}
\includegraphics{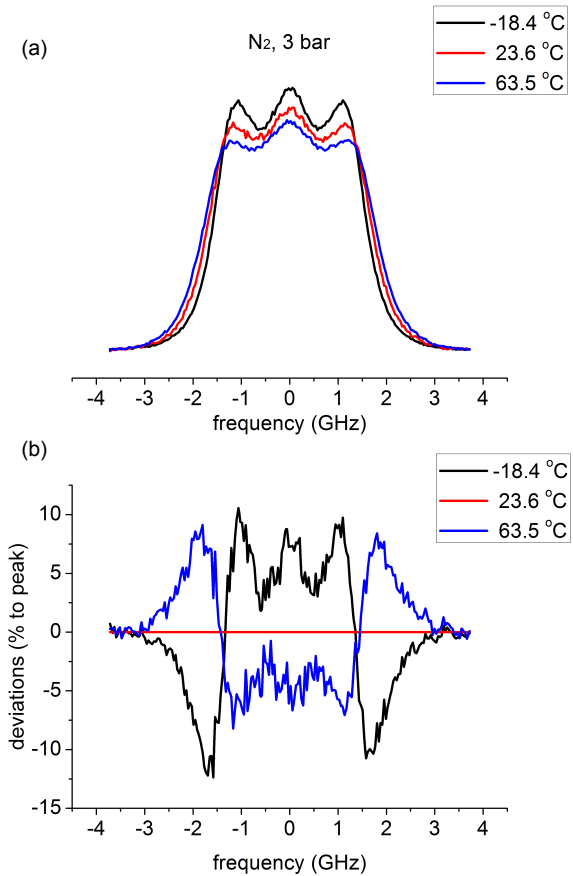}
\caption{\label{fig:N2_3bar_Temperature} (a) The normalized Rayleigh-Brillouin scattering profiles of N$_2$ measured at three different temperatures: $-$18.4$^{\circ}$C (black), 23.6$^{\circ}$C (red) and 63.5$^{\circ}$C (blue). The number density $N$ of the molecules inside the RB scattering cell is kept the same for all three measurements; it corresponds to a pressure of 3 bar at 23.6$^{\circ}$C. (b) The deviations (in percentage of maximum amplitude of the scattering profile) of the scattering profiles at the other two temperatures from that at 23.6$^{\circ}$C.}
\end{figure}

To demonstrate the capabilities of the instrument to record RB-scattering profiles as a function of the gas temperature, measurements were performed for molecular nitrogen gas at three temperatures: $-$18.4$^{\circ}$C, 23.6$^{\circ}$C and 63.5$^{\circ}$C. The measurements pertain to pressures of approximately 3 bar; the molecular number density was kept constant by following a procedure of filling the cell at 3 bar at room temperature, and subsequently lowering or increasing the temperature while keeping the cell sealed; the pressures are then, 2560, 3000 and 3400 mbar respectively, calculated from the ideal gas law.

To avoid ice condensation on the windows of the light scattering cell at temperatures below the freezing point, significantly reducing the enhancement of the cavity and resulting in large amounts of stray light, the cell is placed inside an isolated box with a flush of dry N$_2$. The resulting RB-profiles are normalized to area unity and compared in Fig.~\ref{fig:N2_3bar_Temperature}, indicating a shift of the Brillouin side peaks from the center as temperature increases. The central Rayleigh peak is broadened due to temperature-dependent Doppler Broadening.

Figure~\ref{fig:N2_3bar_Temperature}(b) displays the differential temperature effects, whereby the experimental RB-profiles recorded at elevated temperature (63.5$^{\circ}$C) and lower temperature (-18.4$^{\circ}$C) are plotted, after subtraction of the profile measured at room temperature. It is demonstrated that the temperature effects in this range result in deviations of about 10\% of the peak amplitudes.

\subsection{SF$_6$ measurements at different pressures}
\begin{figure}
\includegraphics{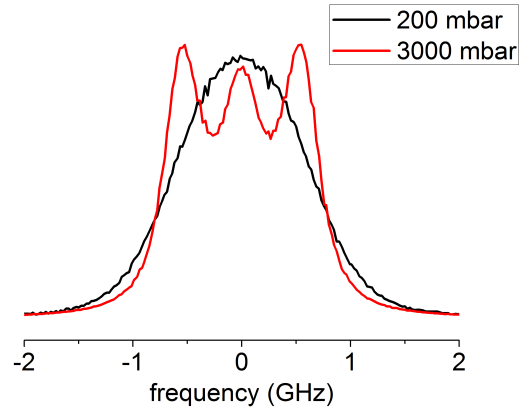}
\caption{\label{fig:SF6_200_and_3000mbar} Normalized scattering profiles of SF$_6$ at 200 mbar (indicated in black) and 3000 mbar (indicated in red). }
\end{figure}

As another application RB scattering profiles were measured in sulphur-hexafluoride (SF$_6$) gas, the molecular species with the largest scattering cross section~\cite{sneep2005}. Figure \ref{fig:SF6_200_and_3000mbar} shows the measured scattering profiles of SF$_6$ at 200 mbar (indicated in black) and 3000 mbar (indicated in red). For 200 mbar, the normalized scattering profile closely resembles a Gaussian shape, resulting from Doppler Broadening. In the normalized spectrum at 3000 mbar the Brillouin side peaks, Stokes and anti-Stokes shifted from the center have become more pronounced than the central Rayleigh peak.  The maximal measured intensity  for 200 mbar are 1900 counts/s, while for 3000 mbar 27,500 counts/s are recorded. At these large amounts of photon counts the scattering profiles are smoothly resolved; even at low pressures the noise level at the peak amplitude is within the 1\% level.

\section{Conclusion}

In this paper, we describe a new instrument for the sensitive spectral measurement of Rayleigh-Brillouin scattering profiles in gases at atmospheric pressures in the UV wavelength range; such profiles, in particular for air and at UV wavelengths, are of importance for modern spaceborne lidar projects such as the ADM-Aeolus project\cite{ESA-SP-1311}. By application of frequency doubling of a titanium:sapphire laser in connection with an enhancement cavity, 5 Watt of UV-light is available in the scattering volume. The scattering measurements on N$_2$ and SF$_6$ demonstrate that rms measurement noise levels of 1\% can be achieved, even when the scattering opening angle for scattered RB light is as small as 0.7$^{\circ}$. The setup is designed to allow for measuring RB profiles as a function of temperature in the range $-$30$^{\circ}$C - 70$^{\circ}$C.

\begin{acknowledgments}
This work was funded by the European Space Agency (ESA) under contract ESTEC-21396/07/NL/HE-CCN-2. The authors wish to thank B. Witschas (DLR Oberpfaffenhofen), W. van de Water (TU Eindhoven), A. S. Meijer (RU Nijmegen), and A. G. Straume and O. Le Rille (ESA) for fruitful discussions, and J. Bouma (VU) for technical assistance.
\end{acknowledgments}

\providecommand{\noopsort}[1]{}\providecommand{\singleletter}[1]{#1}%

\end{document}